# *A Simulation Approach Paradigm:*
# *An Optimization and Inventory Challenge Case Study*

Heru Susanto[123*], Mohammad Nabil Almunawar[1], Mehmet Sabih Aksoy[3]
and Yong Chee Tuan[1],

[1]FBEPS University of Brunei, Information System Group
[2]The Indonesian Institute of Sciences, Information Security & IT Governance Research Group
[3]King Saud University, Information System Department

susanto.net@gmail.com, heru.susanto@lipi.go.id, nabil.almunawar@ubd.edu.bn

**Abstract:**

The paper presents a simulation on automotive inventory and stock issue, followed by evaluated performance of automotif Sector Company, focused on getting optimum profit from supply and demand balancing. Starting by evaluating and verification of customer's document until car delivered to customer. Simulation method of performance is used to evaluate company activity. excess demand of car by customer, not eligible customer to rented a car, number of customer who served and number of customer who served including the driver**,** the last result is number of optimum demand that match with the stock or supply of car by the company. Finally, board of management should be making decision; the first decision is buy the new car for meet with the demand or second decision is recruit new staff for increasing customer service or customer care.

## 1. INTRODUCTION

Simulation refers to a broad collection of methods and application to mimic the behaviour of real system, usually on a computer with appropriate software. In fact, "simulation" can be extremely general term since the idea applies across many fields, industries, and application (Kelton, 2009) and (Susanto et al, 2011). These days, simulation is more popular and powerful than ever since computer and software are better than ever (Whelar & Msefer, 1996). Gu 2011, Proposed a model of jointly managed inventory simulation for thid part stock mode in the steel market's ERP system. Computer simulation deals with models of system. A *System* is facility or process, either actual or planned, such as (Krisztián Bóna. 2004) and (Pucel, 2008):

1. Manufacturing plant with machine, people, transport devices, conveyor belts, and storage space.
2. A bank or other personal-service operation, with different kinds of customer, servers, and facilities like teller windows, automated teller machines (ATMs), loan desks, and safety deposit boxes.
3. A distribution network of plants, warehouse, and transportation links.
4. An emergency facility in a hospital, including personnel, rooms, equipment, supplies and patient transport.
5. A field service operation for appliance or office equipment, with potential customer scatter across a geographic area, service technicians with different qualification, trucks with different parts and tools, and a central depot and dispatch center.

In this research project we focused on the automotive industry, especially in car rental System Company, as parts of sector which needed to evaluate. It contains many steps and procedures in it, with emphasis on customer service, starting from the arrival of customers to the branch or main office and ended by car delivered to the customer.

## 2. PROBLEMS DESCRIPTION

Fortunately, our problems is how to increases and encourage customer service, customer satisfaction and increase the profit as three of main issue in this research project. Car rental System Company named by SRC would like to control and solved its problem that may be appear during the running system. We faced common problem that appear as follow; excess demand, inventory, maintenance cost, highly number of queuing for receiving service from customer care staff [*Figure 1*]. Several terms of condition are determined in order to assign customer who could rent a car form company. Details of these conditions are:
1. Has own home or money deposit at the bank
2. Has valid ID Card or resident Card
3. Has a billing of electricity and water for their home
4. Has driving license from institution which in charge

However, strategic and appropriate decision is needed to giving positioning (Peterson et al, 2007) between optimum demand (Antoniou et al) that should be satisfied and optimum supply (Nien & Chien, 2007) that has correlation to overhead and maintenance cost (Banks, 1983). The simulations are needed for solve problems, at the end of running of modeling and simulation some information should be delivered to management for consideration of company decision making (Michlmayr, 2002) these parameters are:
1. Number of customer who canceled by condition of excess demand.
2. Number of customer who canceled by condition of not eligible and match with term of requirement for rent a car.
3. Number of customer who ordered and eligible without driver
4. Number of customer who ordered and eligible and including driver on it.

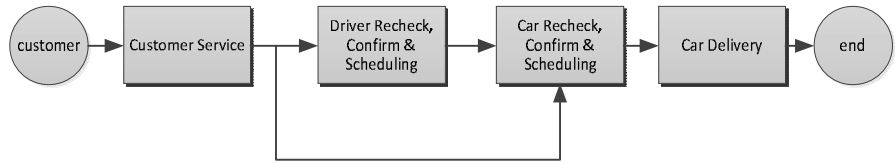

**Figure 1.** Inter department Connection

The overall objectives of this project are to modelling and simulating of service system and car availability to matching with the customer requirement and also for business process reengineering. The Result from this modelling and simulation should be compare with the current running system. In case, current running system has limitation, the new approach and model will be offer to the company, contains of customer service, availability of car for customer demand, especially at the peak season.

## 3. DATA COLLECTION

**Tabel 1:** Overall data

| No | Items | Type | Duration (minutes) | Standard Deviation (minutes) |
|----|-------|------|--------------------|------------------------------|
| 1. | Customer Coming | Exponential | 15 | - |
| 2. | Customer Verification | Normal | 10 | 2 |
| 3. | Survey and The Decision Making | Normal | 30 | 15 |
| 4. | Driver Contract Term | Normal | 5 | 0.01 |
| 5. | Car General Re-Check | Normal | 10 | 0.2 |
| 6. | Engine Re-Check | Normal | 30 | 0.1 |
| 7. | Equipment Re-Check | Uniform | min: 5 | max: 10 | |
| 8. | Insurance Re-Check | Constant | 5 | - |
| 9. | Car's Schedule Re-Check | Constant | 10 | - |
| 10. | Customer Re-scheduling by unavailable car | Normal | 15 | 0.2 |

**Tabel 2:** Rental Prices

| No | Items of Process | Price (US$ / day) | Additional Price (US$ / hour) Max |
|----|------------------|-------------------|-----------------------------------|
| 1. | Car Rental without driver | 40 | 4 |
| 2. | Car Rental with driver | 55 | 6 |

*\* These prices are to all of passenger class of car*

**Tabel 3:** Company Overhead Cost

| No | Items of Process | Price (US$ / day) |
|----|------------------|-------------------|
| 1. | Car Maintenance Fee | 5 |
| 2. | The Available car that not ordered during day | 15 |

## 4. SYSTEM DESCRIPTION

Description of the system is starting by customer arrival, aims to asking or confirm everything regarding to car availability, handled by call centre and customer care department. Four department who deal directly with customers, namely:
1. Customer Service Department
2. Maintenance Department
3. Planning and Scheduling Department
4. Driver Scheduling Department

Each department has process in it, if a department completed with process, than next process will be done by another department, as shown in the picture. The next scenario is, customers who want rent a car with a driver package, high priority, it is meanwhile that customer can directly obtain the car without having to do re-check the car schedule.

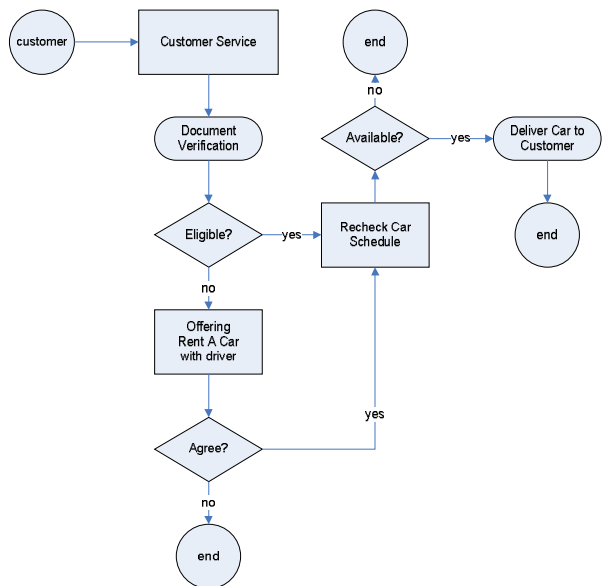

**Figure 2.** Step by step of decision making

**Step1.**
Customers come and submit all documents required to the front office or customer service staff, handling by the customer service department.

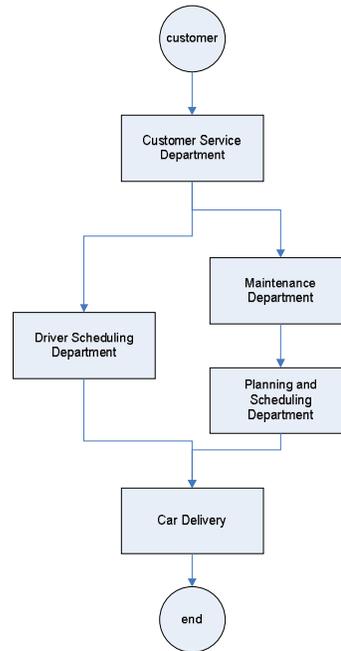

**Figure 3.** Schema of the rental term

At this stage, there will be two outputs; firstly, the customer is considered eligible and qualifies for rent a car; secondly, the customer is not eligible. For second output, its scheme offers car rental with driver from the company, with the assumption that the car will be well controlled during use by the customer. Customers who rent by using the driver will get more priority than others, because security reasons, Figure 3. Scheme for renting a vehicle with a driver is considered to be complete, and the car is ready for delivery to the destination of customer, at this step, required 50 - 65 minutes.

**Step2.**
After the survey and verification is done, the next step is to checking:
1. Engine
2. Equipment Support
3. Insurance

Engine check and equipment checks both these items run as a standard checking, in order to provided safety and comfortable car as company quality service and professional standard. In the insurance check stage, car must be insured for all risk schemas, which covers accident claim, third-party reimbursement for the loss of the vehicle and due to negligence by the customer, Figure 4.

**Step3.**
The next step is checking the schedule of the car, is done by Planning and Scheduling department. Customer is waiting for a confirmation. Schedule verification of the car takes 10 minutes, and follow of the normal statistics. When customer requested a car at certain time, but it does not available for the day concerned, the customer service will offer to reschedule time or if it could not reschedule by customer then the output status is not get the car. For this step takes 10 minutes according to the statistics of constant.

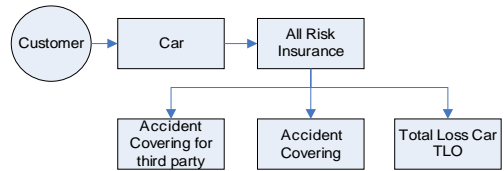

**Figure 4.** Insurance Activity and Covered

**Step4.**

This step is an additional step to the customer who wants to use the driver. Like step 3, when the driver is not available at certain time then company offering reschedule, if approved then the customer will get the driver as well, otherwise customer can not getting car and its driver, Figure 2.

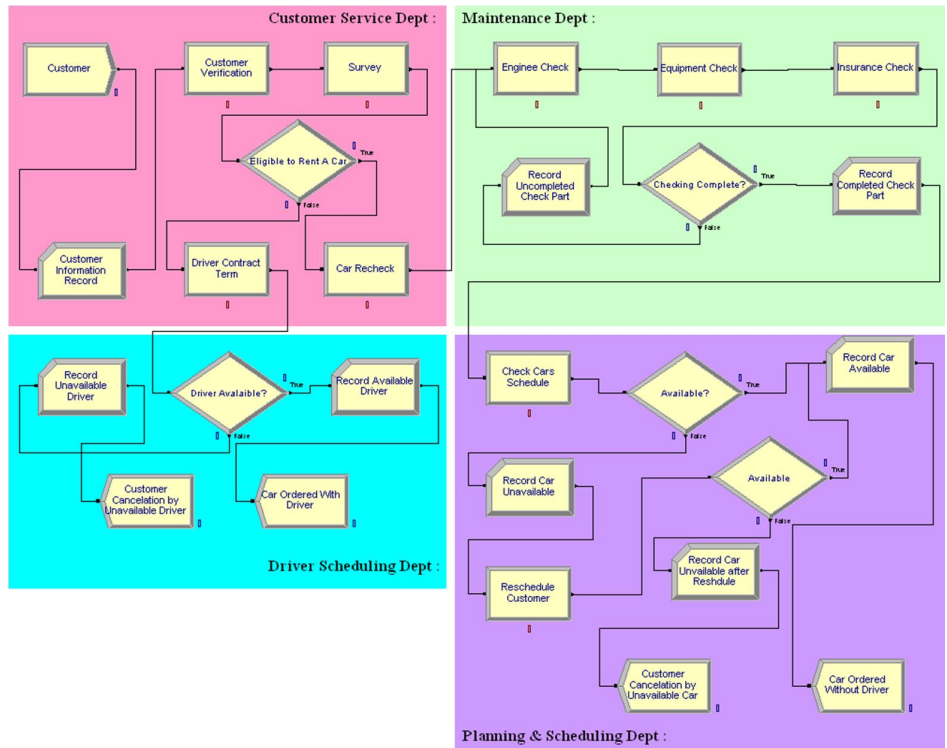

**Figure 5.** Integrated model design and interconnection between department

However, system and model design of the simulation described the relationship between four models according to relevant department of the company are real. Customer service department has 4 processes; *customer verification, survey, offering driver contract term* and *car recheck regarding to it schedule,* as shown in the figure 6. Maintenance department has 3 processes; *Engine Check, Equipment check* and *Insurance check,* output from this department will be as input for Planning and Scheduling department, [*figure 5*]. Planning and Scheduling department has 2 processes; *Check Cars Schedule* and *Reschedule Customer* output from this department will be as output of the system. Two type of output these are car ordered without driver and customer cancelation by unavailable car. Driver scheduling department without process, but contain of two recording activity; recorded available driver and record unavailable driver and two type of outputs; car ordered with driver and customer as shown in the figure 5.

## 5. EXPERIMENTAL RESULT

From the experiment by simulation, there are some facts regarding to demand and supply of car rental company activity. 30 days of repetition of simulation is done. Each day are 12 hours working office, since company opening from 08.00 am – 08.00 pm

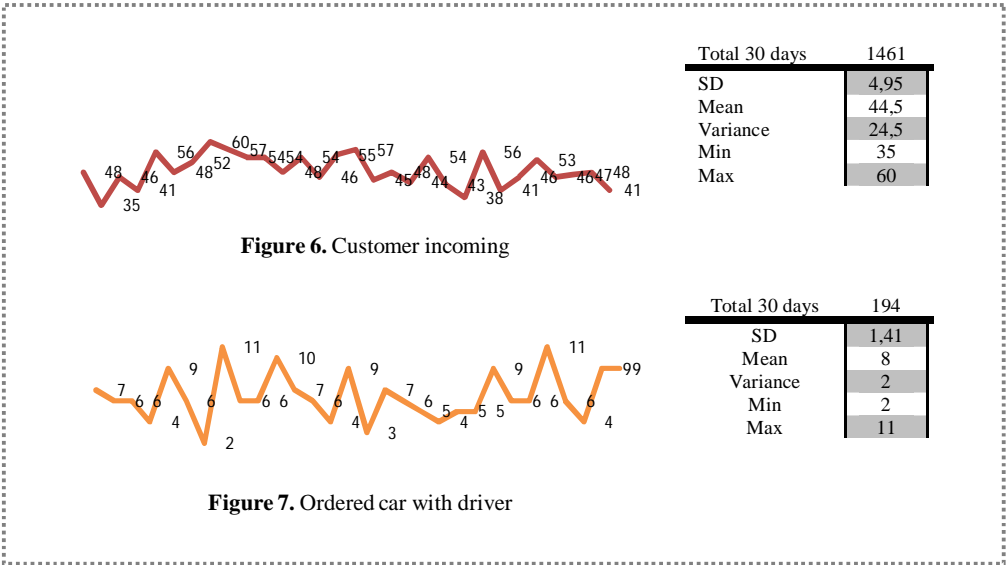

**Figure 6.** Customer incoming

**Figure 7.** Ordered car with driver

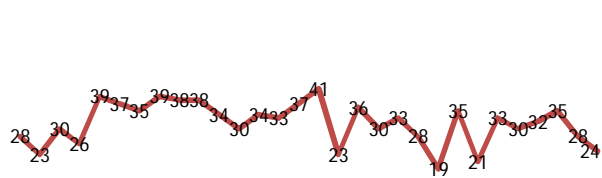

| Total 30 days | 949 |
|---|---|
| SD | 2,83 |
| Mean | 26 |
| Variance | 8 |
| Min | 19 |
| Max | 41 |

**Figure 8.** Ordered Car Without Driver

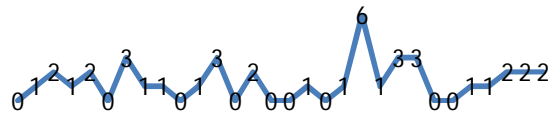

| Total 30 days | 40 |
|---|---|
| SD | 1,41 |
| Mean | 1 |
| Variance | 2 |
| Min | 0 |
| Max | 6 |

**Figure 9.** Unavailable Ordered

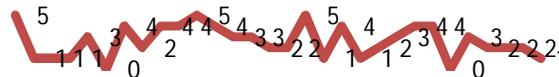

| Total 30 days | 78 |
|---|---|
| SD | 2,83 |
| Mean | 3 |
| Variance | 8 |
| Min | 0 |
| Max | 5 |

**Figure 10.** Unavailable Ordered by Unavailable Driver

## 6. ANALYSIS

**Table 4:** Gap between Customer incoming and Customer Out

| Day to- | Cus. In | Customer Out | | | | Total out | Gap |
|---|---|---|---|---|---|---|---|
| | | Ordered Car with Driver | Ordered Car | Unavailable After Reschedule | Unavailable Driver | | |
| 1 | 48 | 7 | 28 | 0 | 5 | 40 | 8 |
| 2 | 35 | 6 | 23 | 1 | 1 | 31 | 4 |
| 3 | 46 | 6 | 30 | 2 | 1 | 39 | 7 |
| 4 | 41 | 4 | 26 | 1 | 1 | 32 | 9 |
| 5 | 56 | 9 | 39 | 2 | 3 | 53 | 3 |
| ... | ... | ... | ... | ... | ... | ... | ... |
| 27 | 46 | 6 | 32 | 1 | 2 | 41 | 5 |
| 28 | 47 | 4 | 35 | 2 | 2 | 43 | 4 |
| 29 | 48 | 9 | 28 | 2 | 2 | 41 | 7 |
| 30 | 41 | 9 | 24 | 2 | 1 | 36 | 5 |
| Total | | | | | | 1261 | 200 |
| SD | | | | | | | 2,12132 |
| Mean | | | | | | | 6,5 |
| Min | | | | | | | 3 |
| Max | | | | | | | 11 |

**Table 5:** Potential Profit

| Day to- | Ordered Car with Driver (1) | Ordered Car (2) | Revenue (1) | Revenue (2) | total |
|---|---|---|---|---|---|
| 1 | 7 | 28 | 385 | 1.120 | 1.505 |
| 2 | 6 | 23 | 330 | 920 | 1.250 |
| 3 | 6 | 30 | 330 | 1.200 | 1.530 |
| 4 | 4 | 26 | 220 | 1.040 | 1.260 |
| 5 | 9 | 39 | 495 | 1.560 | 2.055 |
| ... | ... | ... | ... | ... | ... |
| 27 | 6 | 32 | 330 | 1.280 | 1.610 |
| 28 | 4 | 35 | 220 | 1.400 | 1.620 |
| 29 | 9 | 28 | 495 | 1.120 | 1.615 |
| 30 | 9 | 24 | 495 | 960 | 1.455 |
| Total | | | | | 48.630 |
| SD | | | | | 35 |
| Mean | | | | | 1.480 |
| Variance | | | | | 1.250 |
| Min | | | | | 1.035 |
| Max | | | | | 2.165 |

**Table 6:** Potential Loss

| Day to- | Unavailable After Reschedule Order | Unavailable Driver | Loss (1) | Loss (2) | Total Loss |
|---|---|---|---|---|---|
| 1 | 0 | 5 | 0 | 200 | 200 |
| 2 | 1 | 1 | 55 | 40 | 95 |
| 3 | 2 | 1 | 110 | 40 | 150 |
| 4 | 1 | 1 | 55 | 40 | 95 |
| 5 | 2 | 3 | 110 | 120 | 230 |
| ... | ... | ... | ... | ... | ... |
| 27 | 1 | 2 | 55 | 80 | 135 |
| 28 | 2 | 2 | 110 | 80 | 190 |
| 29 | 2 | 2 | 110 | 80 | 190 |
| 30 | 2 | 1 | 110 | 40 | 150 |
| Total | | | | | 5.320 |
| SD | | | | | 35 |
| Mean | | | | | 175 |
| Variance | | | | | 1.250 |
| Min | | | | | 0 |
| Max | | | | | 370 |

**Table 7:** Optimum Profit by Optimum Inventory

| Day to- | Inventory (cars) | | | |
|---|---|---|---|---|
| | 30 | 40 | 50 | 53 |
| 1 | 955 | 1.305 | 1.105 | 1.045 |
| 2 | 1.060 | 915 | 715 | 655 |
| 3 | 1.020 | 1.315 | 1.115 | 1.055 |
| 4 | 1.030 | 940 | 740 | 680 |
| 5 | 985 | 1.335 | 1.685 | 1.790 |
| ... | ... | ... | ... | ... |
| 27 | 1.020 | 1.370 | 1.225 | 1.165 |
| 28 | 950 | 1.300 | 1.265 | 1.205 |
| 29 | 1.025 | 1.375 | 1.230 | 1.170 |
| 30 | 1.065 | 1.195 | 995 | 935 |
| Total | 29.690 | 37.440 | 37.160 | 35.745 |
| SD | 78 | 78 | 78 | 78 |
| Mean | 1.010 | 1.250 | 1.050 | 990 |
| Min | 800 | 685 | 485 | 425 |
| Max | 1.140 | 1.490 | 1.795 | 1.900 |

## 7. CONCLUSION REMARKS

Referring to the simulation result and analysis at section 6, company has four alternatives of inventory. Each statistics value related on it mentioned in table 4-7. Some statistics facts are decrypting here; total value, standard deviation, mean / average, maximum and minimum of value. The best recommendation is choice second alternative which is inventory of 40 cars for rent every day. Choosing inventory of 40 cars will give company average of daily profit is US$ 1,250,- and possibility getting lower profit is US$ 685 / day, in other hand this position of inventory would give company maximum profit at position US$ 1,490,- / day. In other hand increasing customer service performance is needed, fortunately, it might be reduce total number of overcoming customer leads to not service customer types and potential loss on gaining profit,

## 8. REFERENCES

1. Constantinos Antoniou, Moshe Ben-Akiva, Michel Bierlaire, and Rabi Mishalani. *Demand Simulation for Dynamic Traffic Assignment*. Massachusetts Institute of Technology, Cambridge, MA 02139.
2. David J. Pucel. *Developing and Implementing Computer Simulated Performance Tests.* 2008. Cambridge University, UK.
3. Gu Y. 2011. *A Model of Jointly Managed Inventory Simulation in the Steel Market's ERP System.* Advanced Material Research. Volume 187 PP 492-497.
4. J. Peterson, J. Larsson, J. Carlsson, P. Andersson. 2007. *Knit on Demand – Simulation of Agile Production and Shop Model for Fashion Products*. ITMC 2007 International Conference.
5. Jerry Banks. 1983. *Discrete System Simulation*. Prentice Hall International Series.
6. John W. Huminel. 1985. *Developments in Business Simulation &Experiential Exercises. A Proposed Interactive Inventory Control Simulation*. University of Vermont.
7. Joseph Whelar & Kamil Msefer. 1996. *Economic Supply & Demand*. MIT Systems Dynamics in Education Project.
8. Kelton W, David. 2009. *Simulation with Arena.* McGraw Hill.
9. Krisztián Bóna. 2004. *Simulation supported optimization of Inventory Control Processes by Application of Genetic Algorithms*. Budapest University of Technology and Economics
10. Nien & Benjamin Chih-Chien. 2007. *Study on Applications of Supply and Demand Theory of Microeconomics and Physics Field Theory to Central Place Theory*. MPRA Paper No. 390.
11. Susanto H, MN Almunawar, YC Tuan, MS Aksoy, WP Syam. 2011. *Integrated Solution Modeling Software: A New Paradigm on Information Security Review and Assessment*. International Journal of Science and Advanced Technology. Volume 1 Issue 10.
12. Von Martin Michlmayr. 2002. *Simulation Theory versus Theory*. Leopold-Franzens-Universität Innsbruck.

## AUTHORS

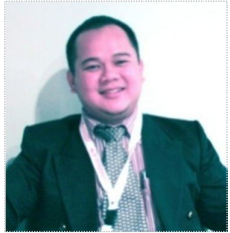 **Heru Susanto** is a researcher at The Indonesian Institute of Sciences, Information Security & IT Governance Research Group, also was working at Prince Muqrin Chair for Information Security Technologies, King Saud University. He received BSc in Computer Science from Bogor Agriculture University, in 1999 and MSc in Computer Science from King Saud University, and nowadays as a PhD Candidate in Information Security System from the University of Brunei.

**Mohammad Nabil Almunawar** is a senior lecturer at Faculty of Business, Economics and Policy Studies, University of Brunei Darussalam. He received master Degree (MSc Computer Science) from the Department of Computer Science, University of Western Ontario, Canada in 1991 and PhD from the University of New South Wales (School of Computer Science and Engineering, UNSW) in 1997. Dr Nabil has published many papers in refereed journals as well as international conferences. He has many years teaching experiences in the area computer and information systems. 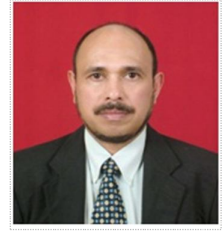

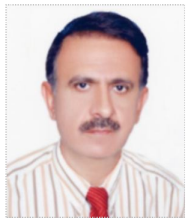 **Mehmet Sabih Aksoy** is a Professor in Information System, King Saud Univrsity. He received BSc from Istanbul Technical University 1982. MSc from Yildiz University Institute of Science 1985 and PhD from University of Wales College of Cardiff Electrical, Electronic and Systems Engineering South Wales UK 1994. Prof Aksoy interest on several area such as Machine learning, Expert system, Knowledge Acquisition, Computer Programming, Data structures and algorithms, Data Mining, Artificial Neural Networks, Computer Vision, Robotics, Automated Visual inspection, Project Management.

**Yong Chee Tuan** is a senior lecturer at Faculty of Business, Economics and Policy Studies, University of Brunei Darussalam, has more than 20 years of experience in IT, HRD, e-gov, environmental management and project management. He received PhD in Computer Science from University of Leeds, UK, in 1994. He was involved in the drafting of the two APEC SME Business Forums Recommendations held in Brunei and Shanghai. He sat in the E-gov Strategic, Policy and Coordinating Group from 2003-2007. He is the vice-chair of the Asia Oceanic Software Park Alliance. 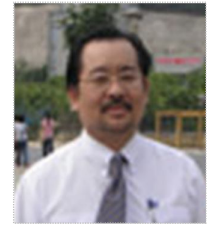